\documentclass{elsarticle}
\usepackage[utf8]{inputenc}
\usepackage{graphicx}
\usepackage{caption}
\usepackage{subcaption}
\usepackage{amsmath}
\usepackage{enumerate}

\begin{document}

\title{The destructive effect of human stupidty: a revision of Cipolla's Fundamental Laws}
\author[fiestin,ib]{Donny R. B\'arcenas}
\author[famaf]{Joel Kuperman}
\author[fiestin,ib]{Marcelo N. Kuperman}
\ead{kuperman@cab.cnea.gov.ar}

\address[fiestin]{Centro At\'{o}mico Bariloche and CONICET, R8402AGP Bariloche, 
Argentina}
\address[ib]{Instituto Balseiro, Universidad Nacional de Cuyo, R8402AGP Bariloche, Argentina}
\address[famaf]{Facultad de Matem\'atica, Astronom\'{\i}a y F\'{\i}sica, Universidad Nacional de C\'ordoba, Ciudad Universitaria, 5000 C\'ordoba, 
Argentina}

\begin{abstract}
In this work we analyze an evolutionary game that incorporates the ideas presented by Cipolla in his work \textit{The fundamental laws of human 
stupidity}. The game considers four strategies, three of them are inherent to the player behavior and can evolve via an imitation dynamics, while 
the fourth one is associated to an eventual behavior that can be adopted by any player at any time with certain probability. This fourth strategy 
corresponds to what Cipolla calls a stupid person. The probability of behaving stupidly acts as a parameter that induces a 
phase transition in the steady the distribution of strategies among the population.
\end{abstract}
\maketitle

\section{Introduction}
In 1988 Cipolla wrote an essay entitled The fundamental laws of human stupidity \cite{cipolla88}. The structure of this essay consisted in several 
chapters with some of them intended for the introduction and discussion of each of the five fundamental laws that according to Cipolla rule the 
human stupidity.

As the concept of stupidity can be ambiguous it is important to properly frame the meaning we embrace here. A stupid  person is someone given to 
unintelligent decisions or acts but here  we consider those acts within a social context. Stupidity should not be understood as the opposite 
of intelligence. In fact, according to some of the ideas of Cipolla, none of us gets rid of or will never get rid of a brief moment of stupidity.

The association  of stupidity to a source of  collective troubles and nuisance and  to the origin of social scourges  has been manifested through 
history in several nowadays popular quotes . Among them it is worth citing a phrase credited to  A. Dumas: "One thing that humbles me 
deeply is to see that human genius has its limits while human stupidity does not" \cite{dumas}. B. Russell, in his essay \textit{The triumph of 
stupidity} wrote:  "The fundamental cause of the trouble is that in the modern world the stupid are cocksure while the intelligent are full of doubt" 
\cite{russ}.

Cipolla starts by preventing us about the silent danger of stupidity in the first law. There he affirms that detecting a stupid person is a hard 
task. This law states  that always and inevitably everyone underestimates the number of stupid individuals in circulation.

While it could be tempting to associate stupidity with lack of education or training,  Cipolla affirms that the probability that a certain 
person be stupid is independent of any other characteristic of that person. This is the content of the second law. This somehow suggests that there 
is a natural stupidity, which resists any  academic training. 

But again, we are here interested in an operational definition of stupidity, the  one that is dangerous for others, an idea that can be summarized 
by a quote taken from one of  M. Atwood novels, \cite{atwood}: “Stupidity is the same as evil if you judge by the results.” 

This idea is expressed in Cipolla's third law: A stupid person is a person who causes losses to another person or to a group of persons while himself 
deriving no gain and even possibly incurring losses.

This law also suggest the definition of  three other phenotypes that complement the stupid group (S). These three groups, according to Cipolla,  are 
the intelligent people (I), whose actions benefit both themselves and others, the bandits (B), who benefits themselves at the expense of others, and 
finally the  helpless or unaware people (U), whose actions enrich others at their own expense.

Stupid people are dangerous and damaging because their behavior is hard to understand and predict from a rational point of view. The  bandit’s 
actions, while producing some damage, obey a predictable pattern of rationality. The possibility to foresee the behavior of a bandit can help an 
individual to build up defenses against it. On the contrary, when facing a stupid person this is  impossible.
So, while most of the time the evil has a clear face and is easily identifiable, stupidity is not. This biased evaluation is what is considered in 
the forth law: Non-stupid people always underestimate the damaging power of stupid individuals. In particular non-stupid people constantly forget 
that at all times and places and under any circumstances to deal and/or associate with stupid people always turns out to be a costly mistake.

The effect of stupidity and the difficulty to recognize it is what leads us to the fifth law: A stupid person is the most dangerous type of person.

Cipolla characterized the four groups in terms of two parameters; the own gains or losses
$p$, and the gains or  losses that an individual inflicts on others, $q$. The payoff resulting from the interaction between two persons can be 
definedin terms of  these quantities associated to the identification of the participants with one of the four defined groups. 
These four groups can then be characterized by the range of values adopted  by  $p$ and $q$ as follows :
\begin{itemize}
\item[S] : $p_s \leq 0$ y $q_s < 0$
\item[U]: $p_u \leq 0$ y $q_u\geq 0$
\item[I]: $p_i > 0$ y $q_e \geq 0$
\item[B]: $p_b > 0$ y $q_b < 0$
\end{itemize}

In \cite{joel} we presented a four strategy game based on this four groups. The payoff of the strategies was defined by the values of $p$ and $q$ 
as presented in Table \ref{tabla1}, that indicates which is the payoff of the strategy at the file when competing with the strategy at the column
\begin{table}[h]
\centering
\begin{tabular}{c| c| c|c |c}
 &{\bf S} & {\bf U} &{\bf I} & {\bf B}\\
\hline
{\bf S} & $p_s +q_s$ & $p_s+q_u$& $p_s+q_i$ & $p_s+q_b$  \\
\hline
{\bf U} & $p_u +q_s$ & $p_u +q_u$& $p_u +q_i$ & $p_u +q_b$  \\
\hline
{\bf I} & $p_i +q_s$ & $p_i +q_u$& $p_i +q_i$ & $p_i +q_b$  \\
\hline
{\bf B} & $p_b +q_s$ & $p_b +q_u$ & $p_b +q_i$ & $p_b +q_b$
\end{tabular}
\caption{Payoff Table} \label{tabla1}
\end{table}

The results shown in \cite{joel} supported the validity of the law enunciated by Cipolla that contained some appreciations about the dangerousness 
of the presence of stupid people. In this work, the evaluation of the effect of the actions of this group was done by calculating  
the total wealth of the population in the steady state of the strategy profile resulting from an evolutionary game and an imitation dynamics.
The presence of stupid people not only undermined the total wealth but also promoted the inhibition of cooperative behaviors represented by 
intelligent and unaware people. In that work we could not find a critical value for the fraction of stupids that could divide the behavior of the 
system into two different regimes. According to the the first law it is not possible to know the number of stupid individuals in a 
population, so it might be interesting to find different regimes for different fraction of stupid people.

In the present work we have adopted a different approach, interpreting Cipolla' s idea in a different way.
We will consider a three strategy game, where the participating strategies will be (I), (B) and (U) and we will let any individual to occasionally 
behave as a stupid person with a given probability. This probability is the parameter that will govern the amount of stupid people at each time.

This means that being stupid will not be a permanent state by an occasional state accounting for the possibility that at any time any individual can 
behave stupidly. In the following sections we present a more thorough description of the the model and the numerical results.

\section{The model}

As mentioned before,  we are going to consider an evolutionary game, with four strategies though one of them, the (S), differs from the 
others in the 
sense that it is not durable and it can not be imitated. Any individual can be stupid at any time and during one time step with probability $\rho_s$ .

While stupidity will not be a permanent condition on this version of the game, we still need to define the payoff of each of the three strategies 
when playing between them and when ocassionally confronting with a stupid person. We also need to define the payoff that 
an eventual stupid player may receive.
We introduce then a 4x4  payoff matrix
\[A=\begin{pmatrix}
p_s +q_s & p_s+q_u& p_s+q_i & p_s+q_b\\
p_u +q_s & p_u +q_u& p_u +q_i & p_u +q_b \\
p_i +q_s & p_i +q_u& p_i +q_i & p_i +q_b\\
p_b +q_s & p_b +q_u & p_b +q_i & p_b +q_b
\end{pmatrix}\]

To compare the effects of this new model with those obtained in \cite{joel}, we will adopt the  values in Table \ref{tabla2}.

\begin{table}[h]
\centering
\begin{tabular}{| c| c|c |c|c|c|c|c|}
\hline
 $x_i$ &$x_b$ &$x_d$ &$x_e$&$y_i$&$y_b$&$y_d$&$y_e$\\
\hline
 1 &[1.1,2] &[-2,-1] &[-2,-1]&1&-1&1&-1\\
 \hline
\end{tabular}
\caption{Chosen values for the payoff matrix} \label{tabla2}
\end{table}
The only Nash equilibrium of this game is strategy (B). As shown in \cite{joel}, considering the chosen values for the payoffs, the sub game in which 
only the strategies  (I) and (B) participate constitutes  a Prisoner's Dilemma (PD) or a Donation Game \cite{sigm10} .

Despite that there is only one Nash equilibrium and thus an evolutionary game ruled by the replicator mean field equations will have only one steady 
state, when considering a spatially extended game with players located on top a network, the steady state can show a different steady configuration 
\cite{dur,kup3,yukov,pavlogiannis}.

\begin{figure}[h]
\includegraphics[width=.8\textwidth]{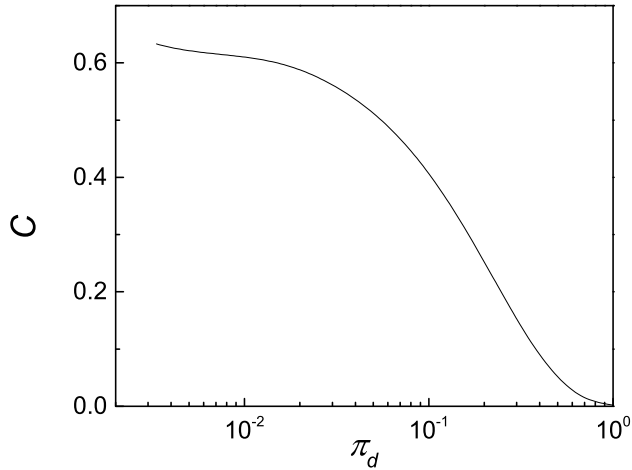}
    \caption{Mean clustering coefficient of the used networks as a function of the degree of disorder}
  \label{clust}
\end{figure}

In order to compare the results obtained here with those previously shown in \cite{joel} we locate the players on top of the networks used in that 
work . We consider a particular family of regular networks, i.e. with all the node having the same degree, but  with a tunable degree 
of disorder. These networks. described in \cite{kup2} present a topology that  varies according to a disorder parameter $\pi_d$. This parameter is 
responsible for the change in the value of the clustering coefficient $C$ of the network.
This dependence is shown if Fig. (\ref{clust})

We will adopt a simple evolutionary dynamics for the strategies considering a deterministic imitation. In each round a selected player plays with all 
its neighbors. In turn, these  neighbors  do the same with their own. After that selected player analyses its performance or earnings 
and compares them with that of its neighbors. Then, it adopts the strategy of the player with the highest gain, that eventually can be its own one. 
In case of tie the choice  is decided at random. This update dynamics is the simplest one, representing a deterministic imitation and closely linked 
to the replicator dynamics \cite{hofb}. 

Players will play according to their chosen strategy, that can be (I), (B) or (U) but at any time, any player can adopt the strategy (S) with 
probability $\rho_s$. This election will not be permanent, will last only one time step (there is always a probability  $\rho_s^n$ of 
adopting the (S) strategy during $n$ consecutive steps) and after that, the player will adopt the original behavior or change it to imitate the 
neighbor with the highest payoff. We recall that the (S) 
strategy can not be imitated, but this is not an issue as under any circumstance a player adopting the (S) strategy will obtain the higher payoff.
The fact that there is  probability $\rho_s$ of adopting strategy (S) means that at each time step there is a mean effective population of $\rho_s N$ 
stupids, where $N$ is the total population size.

\section{Results}

\begin{figure}
\centering
\begin{subfigure}{.5\textwidth}
  \centering
  \includegraphics[width=.8\linewidth]{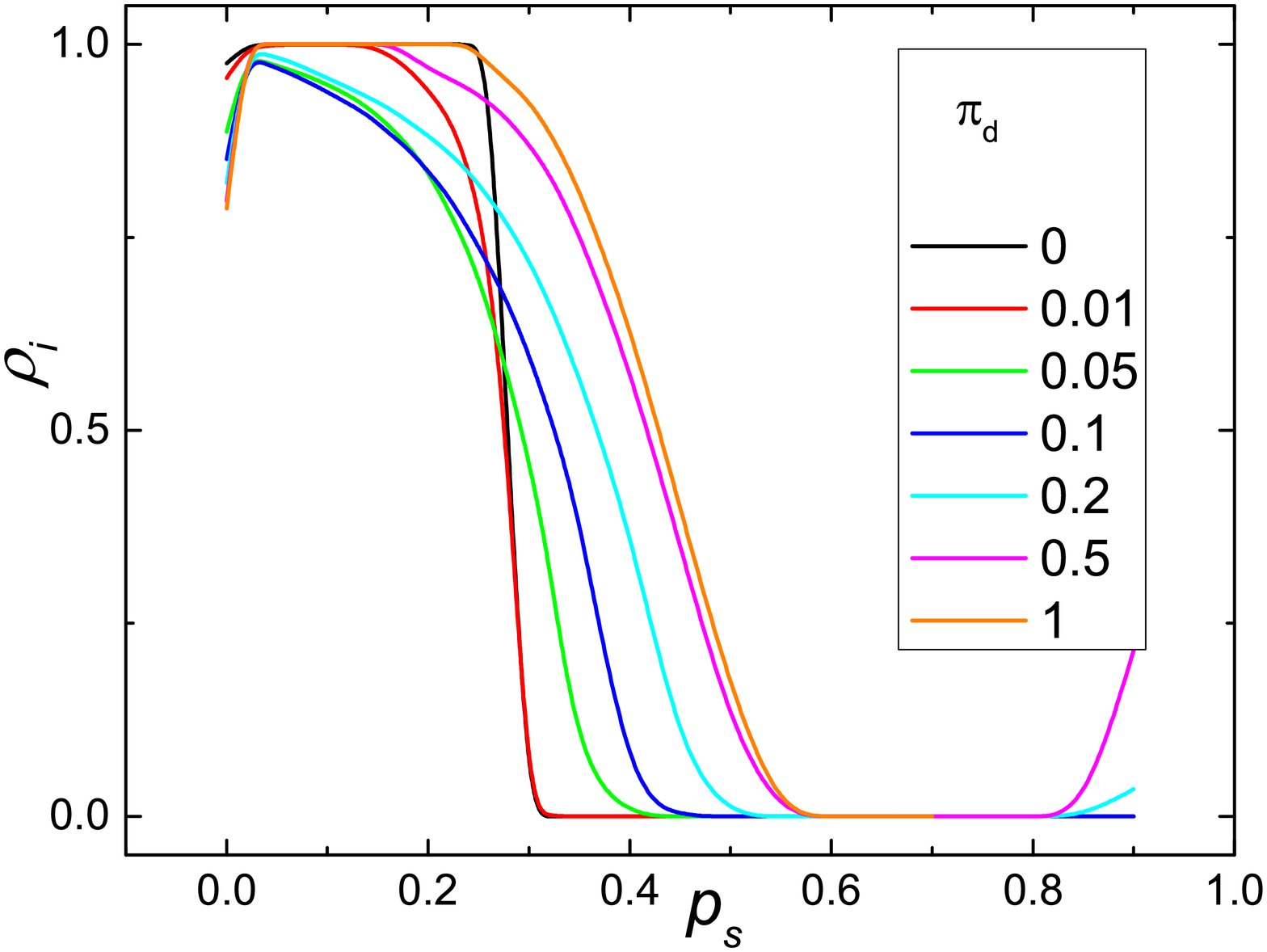}
  \label{fig:sub1}
\end{subfigure}%
\begin{subfigure}{.5\textwidth}
  \centering
  \includegraphics[width=.8\linewidth]{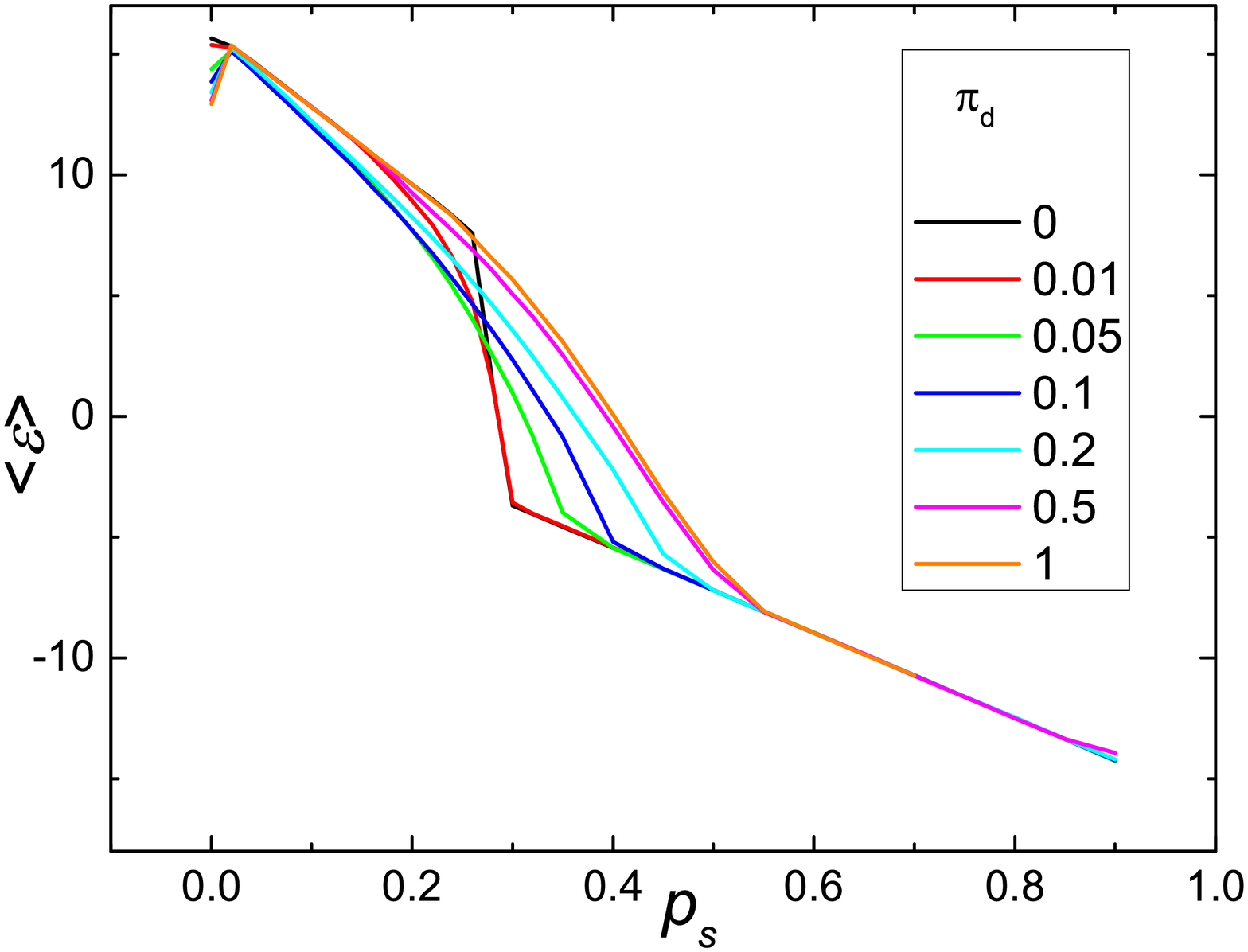}
  
  \label{fig:sub2}
\end{subfigure}
\caption{This plots shows the fraction of (I) individuals $\rho_i$ (left) and the main gain of the population in one round $<\epsilon>$ (right) as a 
function of $\rho_s$. All the curves correspond to a the steady state}
\label{figura1}
\end{figure}

In our simulations we considered networks of $N=10^5$ nodes and the system evolved untill reaching a steady state. The degree of disorder $\pi_d$ and 
the probability of adopting the (S) strategy at each 
time step $\rho_s$ were chosen as parameters.  
The results are shown in Fig. (\ref{figura1})

We observe that the topology of the network has a minor effect on the evolution of the strategy profile of the population. On the contrary
the probability of adopting the (S) strategy plays a crucial role. The system shows the existence of a critical value of $\rho_s$ separating the 
evolution of the system towards two differently regimes. For low values of $\rho_s$ the dominating strategy is (I) while for higher values, (B) 
dominates.
This is clearly reflected in the left panel of Fig (\ref{figura1}) showing the fraction of (I) in the steady state and also in the right panel 
showing the mean gain of the population. 
Clearly, 
the prevalence of (B) attempts against the wealth of the population, and this prevalence is promoted by the sporadic appearance of the (S) behavior.
The interesting additional feature observe in this work is the existence of two well defined ranges, and a critical $\rho_s$ values more defined when 
the network is more ordered.

The presence of a sharp transition between a cooperative to a defective steady state was not observed in \cite{joel}.
To understand the relevance of this result we refer to previous results involving evolutionary games. There are several examples of the 
effect of locating the players on networks with different topologies \cite{dur,kup3,kup2,now0,now3,sza2,now1,kup1,roc,ass,ift,vuk2,gan}.
These works show that the evolutionary behaviour of the strategies  might be affected by the underlying topology of links 
between players, sometime promoting cooperative states even when the Nash equilibrium is the defective strategy.
In these examples, the topology of the network is the factor responsible for different regimes. Here we show that while the topology of the networks 
plays a non negligible role, the most important parameter is the probability of a player to adopt the (S) strategy.

\begin{figure}[!tbp]
\includegraphics[width=.8\textwidth]{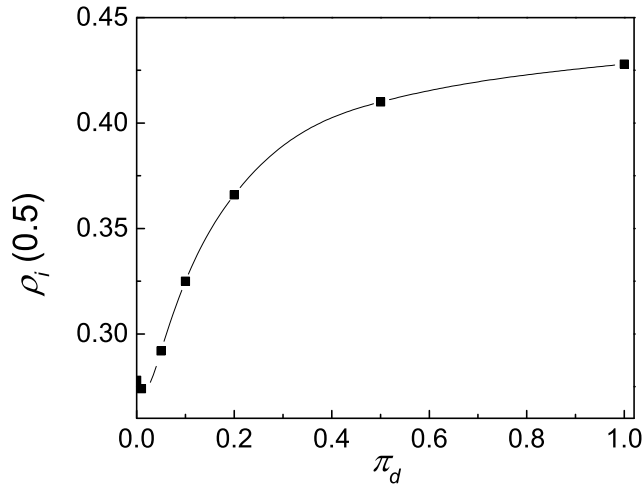}
    \caption{This plot shows the value of $\rho_s$ at which $\rho_i=0.5$ as a function of $\pi_d$. The curve is a spline for helping visualization.}
  \label{icrit}
\end{figure}

The results show that as $\rho_i$ increases there is a transition from a scenario where (I) is prevalent to another one where most of the population 
behaves as (B). The transition is sharper for highly ordered network and turns smoother as $\pi_d$ increases. Also, the critical value at which this 
transition occurs moves to right with increasing $\pi_d$. This fact is shown in Fig, (\ref{icrit})

\section{Conclusions}

As we stated in the introduction, the work by Cipolla should be understood in a cartoonish way. Nevertheless, it implies certain facts that deserve 
to be taken into consideration. It is in this spirit that we analyzed the work by Cipolla and the results that can be obtained by translating these 
ideas into a 
mathematical model. Lets start by the first law. It affirms that we always  underestimate the number of stupid individuals in 
circulation. Inspired by this statement we wander whether despite the density of stupids is impossible to calculate there is a critical density 
separating two different scenarios or not. We mean by this to evaluate the possibility that only by reaching a threshold density, the group of stupid 
people can inflict a considerable harm to the entire population.
For that we proposed a model in which any individual is susceptible of behaving stupidly at any time and with a certain probability.
Our results show the existence of a critical probability $\rho_s \approx 0.35$. The transition from the cooperative regime to the defective one is 
only sharp  for slightly disordered networks, turning smoother as the disorder increases. Also, the increasing disorder produces a displacement of 
$\rho_s$ to higher values. The curves show that while for low values of $\rho_s$ the disorder attempts against (I), the situation is reversed for 
higher degrees of disorder. These results agree with those obtained in \cite{joel} for the case when the fraction of (I) individuals remains constant 
along the whole simulation.
Another interesting feature is the increase in the fraction of (I) for higher values of $\rho_s$. This phenomenon can be attributed to  a screening 
effect played by the (S) population. The (I) players can not be affected and tempted by the presence of (B) ones and then they survive. This effect 
has been also found in \cite{joel}.

\end{document}